\documentclass[aps,prl,twocolumn,showpacs,groupedaddress]{revtex4}

\usepackage{graphicx}
\begin{document}
\title{Berry's phase with quantized field driving:  effects of inter-subsystem coupling}
\author{L. C. Wang, H. T. Cui, and X.X. Yi}
\affiliation{Department of physics, Dalian University of
Technology, Dalian 116024, China}
\date{\today}
\begin{abstract}
The effect of inter-subsystem coupling on the Berry phase of a
composite system as well as that of its subsystem is investigated
in this paper. We analyze two coupled spin-$\frac 1 2 $ with one
driven by a quantized field as an example, the pure state
geometric phase of the composite system as well as the mixed state
geometric phase for the subsystem is calculated and discussed.
\end{abstract}
 \maketitle

The concept of geometric phase was first introduced by
Pancharatnam \cite{pancharatnam} in his study on the interference
of light in distinct states of polarization, and later was
extended to its quantal counterpart by Berry \cite{berry84}, who
shown that the state of a quantum system acquires a purely
geometric feature in addition to the usual dynamical phase when it
is varied slowly and eventually brought back to its initial form.
The geometric phase has been extensively studied
\cite{shapere89,thouless83,sun90} and generalized, for example to
nonadiabatic evolution \cite{aharonov87}, mixed states
\cite{uhlmann86,sjoqvist00a}, and open systems \cite{carollo03}.
All these studies were based on the semiclassical theory in the
sense that the driving field itself has never been quantized. The
effect due to the field quantization on the geometric phase was
theoretically studied \cite{fuentes02} through a cavity QED model,
where the cavity mode acted as the driving field, this work
appears as the first study on the Berry phase with quantum field
driven since no research devoted to this problem before. The other
direction to which the Berry phase have been generalized is the
Berry phase in composite systems\cite{yi03}, this study was
motivated by the application of geometric phase in quantum
information  processing as well as the theory of geometric phase
itself.

From the application aspect, the implementation of quantum
information by geometric means hold the merit of some built-in
fault-tolerant features, because most systems for this purpose are
composite, the study on the Berry phase of composite systems is
highly required. From another aspect, since the geometric phase of
entangled systems have attracted interest for its connection to
the topology of the SO(3) rotation group \cite{milman03} and
Bell's theorem \cite{bertlmann03}, and the  entanglement may be
created only via interactions or joint measurements, how
inter-subsystem coupling may affect the geometric phase of a
composite system is  of interest then.

In this Letter, we investigate the behavior of the geometric phase
of a bipartite system with inter-subsystem coupling, one of the
subsystems is driven by a quantized single mode of field, we
examine the effect of inter-subsystem coupling on the pure state
geometric phase of the composite system, as well as on the mixed
state geometric phase of the subsystem. The composite system is
designed to undergo an adiabatic and cyclic evolution while the
subsystems remain on their non-transition state \cite{yi04}. An
example of two coupled spin-$\frac 1 2 $ systems with one driven
by a quantized  mode of field is presented to detail the
representation. We calculate and analyze the effect of spin-spin
coupling on the geometric phase of the composite system and those
of the subsystem. The results presented in this Letter are
twofold; it is an extension of the mixed state geometric phase to
the case with quantized field driving, and it would generalize the
study on the Berry phase induced by vacuum to composite systems
with inter-subsystem couplings.

Consider a composite system consisting of two interacting
spin-$\frac 1 2 $ subsystems in the presence of a single quantized
mode of field, in the rotating wave approximation (RWA) the
Hamiltonian governing such a system reads
\begin{equation}
H=\frac{\omega}{2}(\sigma_1^z+\sigma_2^z)+\nu
a^{\dagger}a+\lambda(\sigma_1^+ a+
\sigma_1^-a^{\dagger})+J\sigma_1^z\sigma_2^z, \label{hamil1}
\end{equation}
where $\omega$ is the transition frequency between the eigenstates
of the spin-$\frac 1 2 $, which was assumed to be the same for the
two subsystems, $\nu$ is the frequency of the field described in
terms of the creation and annihilation operators $a^{\dagger}$ and
$a$, respectively, $\lambda$ stands for the coupling constant
between the field and the subsystem 1, and $J$ describes the
coupling constant between the two spin-$\frac 1 2 $. This model
also can be understood to describe two two-level atoms with
dipole-dipole interactions, one of the atoms interacting with a
quantized cavity field, the difference is  the interaction
$J\sigma_1^z\sigma_2^z$ is not a typical dipole-dipole coupling,
but rather a toy model describing a spin-spin interaction,
nevertheless, the presentation in this Letter can be generalized
to the cavity QED system with dipole-dipole coupling, in which the
observation of such an effect is feasible with current technology.

To proceed further, let us first recount the
method\cite{fuentes02} for calculating the geometric phases in the
full quantized regime. In the standard semiclassical theory, the
field operators $a^{\dagger}$ and $a$ are replaced by the
classical amplitude with rotation factors $e^{i\phi}$ and
$e^{-i\phi}$, respectively. Changing $\phi$ slowly from $0$ to
$2\pi$, the system would transport round a circuit in the
parameter space and acquire a geometric phase in addition to the
familiar dynamical phase factor. In the fully quantized context,
the same procedure to generate an analogous phase change in the
state of the field is needed, in practise the phase shift operator
$U(\phi)=e^{-i\phi a^{\dagger}a}$ applied adiabatically to the
Hamiltonian of the system may meet this need \cite{fuentes02},
this would give rise to the following eigenstate of the
Hamiltonian Eq.(\ref{hamil1}) after the phase shift operator
applied
\begin{widetext}
\begin{eqnarray}
|\psi_1\rangle&=& \cos\frac{\alpha}{2}e^{-in\phi}|e_1,e_2,n\rangle
+\sin\frac{\alpha}{2}e^{-i(n+1)\phi}|g_1,e_2,n+1\rangle,\nonumber\\
|\psi_2\rangle&=&
-\sin\frac{\alpha}{2}e^{-in\phi}|e_1,e_2,n\rangle
+\cos\frac{\alpha}{2}e^{-i(n+1)\phi}|g_1,e_2,n+1\rangle,\nonumber\\
|\psi_3\rangle&=& \cos\frac{\beta}{2}e^{-in\phi}|e_1,g_2,n\rangle
+\sin\frac{\beta}{2}e^{-i(n+1)\phi}|g_1,g_2,n+1\rangle,\nonumber\\
|\psi_4\rangle&=& -\sin\frac{\beta}{2}e^{-in\phi}|e_1,g_2,n\rangle
+\cos\frac{\beta}{2}e^{-i(n+1)\phi}|g_1,g_2,n+1\rangle,
\label{eigens}
\end{eqnarray}
\end{widetext}
with the corresponding eigenvalues, $
E_{1,2}=\frac{\omega+(2n+1)\nu}{2}\pm
\frac{2}{\sqrt{4\lambda^2(n+1)+(\omega+2J-\nu)^2}},$ $
E_{3,4}=\frac{(2n+1)\nu-\omega}{2}\pm
\frac{2}{\sqrt{4\lambda^2(n+1)+(\omega-2J-\nu)^2}},$ where
$\alpha$ and $\beta$ are defined as
\begin{eqnarray}
\cos\alpha &=&
\frac{\omega+2J-\nu}{\sqrt{4\lambda^2(n+1)+(\omega+2J-\nu)^2}},\nonumber\\
\cos\beta &=&
\frac{\omega-2J-\nu}{\sqrt{4\lambda^2(n+1)+(\omega-2J-\nu)^2}},
\label{alpha}
\end{eqnarray}
and $|x_1,y_2,n\rangle\equiv
|x_1\rangle\otimes|y_2\rangle\otimes|n\rangle (x,y=g,e)$
represents the basis of the composite system. In the same manner
as that in the standard semiclassical theory, the Berry phase are
calculated as
$\gamma_i=i\int_0^{2\pi}\langle\psi_i|\frac{\partial}{\partial
\phi}|\psi_i\rangle d\phi$, it yields
\begin{eqnarray}
\gamma_1&=&2n\pi+\pi(1-\cos\alpha),\nonumber\\
\gamma_2&=&2(n+1)\pi-\pi(1-\cos\alpha),\nonumber\\
\gamma_3&=&2n\pi+\pi(1-\cos\beta),\nonumber\\
\gamma_2&=&2(n+1)\pi-\pi(1-\cos\beta),\label{berryp1}
\end{eqnarray}
As shown in Ref.\cite{fuentes02}, the Berry phase are different
from zero even for the driving field in the vacuum state ($n=0$),
this indicates that the vacuum field may introduce a correction in
the Berry phase. Furthermore, the Berry phases Eq. (\ref{berryp1})
would return to the semiclassical results when the driving field
are prepared in a coherent state with large mean photon number,
for more detail, we refer the reader to Ref.\cite{fuentes02}, here
we mainly focus on the effects due to the inter-subsystem
coupling. From Eq.(\ref{berryp1}), the effects due to the
inter-subsystem coupling are obvious, all Berry phases tend to
zero (or $2m\pi$, $m$ an integer) with the coupling constant
$J\rightarrow \infty$ compared to $\lambda\sqrt{(n+1)}$, this fact
makes the observation of the vacuum induced Berry phase in the
composite system difficult, in other words, to observe the Berry
phase induced by the vacuum field in the composite system, the
system with small inter-subsystem coupling relative  to the
particle-field interaction is required.  The composite system
would acquire geometric phase $\pi$ or $-\pi$ when there is no
inter-subsystem coupling and with resonant field-particle coupling
($\cos\alpha=\cos\beta=0$) , it nevertheless is not the case when
the inter-subsystem couplings take place, the state would acquire
a Berry phase different from $\pm \pi$ even if the particle-field
coupling is on resonance ($\cos\alpha=-\cos\beta\neq 0$); this
point can be understood as follows, the states of the subsystem 2
does not change during the interaction, hence the inter-subsystem
coupling only results in a level shift to the subsystem 1, this
leads to the effect different from the case without
inter-subsystem couplings. Mathematically, the effect of the
inter-subsystem coupling in Berry's phase of the composite system
may be simulated by the detuning $\delta=\omega-\nu$ in this
model, however, when we extend this representation to a system
with intra-variable coupling, the result will be changed, we will
mention it again later on.

The physical meaning of the term $2\pi m$ ($m$, an integer) in
Eq.(\ref{berryp1}) can be exhibited by the same scheme as in
\cite{fuentes02}, i.e., by introducing the second mode of the
field with creation and annihilation operators $b^{\dagger}$ and
$b$ which initially does not interact with the two spin-$\frac 1 2
$ nor the first mode of the field. The Hamiltonian describing such
a system has the form
\begin{equation}
H_0^{2q}=\frac{\omega}{2}(\sigma_1^z+\sigma_2^z)+\nu
a^{\dagger}a+\nu b^{\dagger}b+\lambda(\sigma_1^+ a+
\sigma_1^-a^{\dagger})+J\sigma_1^z\sigma_2^z, \label{hamil2}
\end{equation}
where the second mode with the same frequency as the first one was
assumed. The eigenstates of this Hamiltonian are
\begin{equation}
|\psi_i^{2q}\rangle=|\psi_i^0\rangle\otimes|n^{'}\rangle, i=1,
...,4 \label{neigens}
\end{equation}
with $|\psi_i^0\rangle$ representing the eigenstates of the
Hamiltonian Eq. (\ref{hamil1}). The state vector is a product
state of the Fock state $|n^{'}\rangle$ of the second mode of the
field and the two spin-$\frac 1 2 $ particles with one driven by
the first field and with spin-spin couplings
$J\sigma_1^z\sigma_2^z$. We proceeded to calculate the Berry phase
of the system by changing adiabatically the Hamiltonian
Eq.(\ref{hamil2}) via the unitary transformation
\begin{equation}
U(\theta,\phi)=exp(-i\phi J_z)exp(-i\theta J_y)
\end{equation}
with $J_z=\frac 1 2 (a^{\dagger}a-b^{\dagger}b)$ and $J_y=\frac
{1}{2i}(a^{\dagger}b-ab^{\dagger})$, where $\theta$ and $\phi$
stand for slowly varying parameters. The transformed Hamiltonian
$H^{2q}=U(\theta,\phi)H_0^{2q}U^{\dagger}(\theta,\phi)$ describes
two coupling spin-$\frac 1 2 $ interacting simultaneously with the
two modes of the field. The four eigenstates Eq.(\ref{neigens})
acquires geometric phase when $\phi$ is altered from $0$ to $2\pi$
adiabatically as
\begin{eqnarray}
\gamma_1^{2q}&=&\pi\cos\theta(n-n^{'}+\frac 1 2)
-\frac{\pi}{2}\cos\theta\cos\alpha,\nonumber\\
\gamma_2^{2q}&=&\pi\cos\theta(n-n^{'}+\frac 1 2)
+\frac{\pi}{2}\cos\theta\cos\alpha,\nonumber\\
\gamma_3^{2q}&=&\pi\cos\theta(n-n^{'}+\frac 1 2)
-\frac{\pi}{2}\cos\theta\cos\beta,\nonumber\\
\gamma_4^{2q}&=&\pi\cos\theta(n-n^{'}+\frac 1 2)
+\frac{\pi}{2}\cos\theta\cos\beta.\label{bpt1}
\end{eqnarray}
The contribution $\pi\cos\theta(n-n^{'})$ may be understood as a
phase acquired of a polarized field whose polarization slowly
rotates and performs a closed loop in the poincar\'{e}'s sphere
\cite{fuentes02},  hence it is inter-subsystem coupling
independent. Terms $\frac {\pi} {2} \cos \theta(1\pm\cos\xi),
\xi=\alpha, \beta$ are pure effects of field quantization which
has no semiclassical correspondence, it makes nonzero contribution
even if the particle-field coupling is on resonance, this is the
very effect due to the spin-spin couplings.
\begin{figure}
\includegraphics*[width=0.95\columnwidth,
height=0.6\columnwidth]{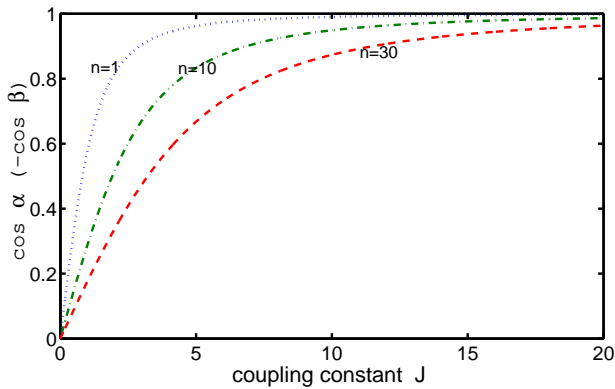} \caption{$\cos\alpha$ (or
$\cos\beta$) as a function of spin-spin coupling constant $J$ (in
units of $\lambda$, the spin-field coupling) with different photon
number $n$. } \label{fig1}
\end{figure}

Now we turn our attention  to study the Berry phase of the
subsystem. Uhlmann was the first to address the issue of mixed
state geometric phase \cite{uhlmann86}, the analysis was
generalized to mixed states undergoing unitary evolution
\cite{sjoqvist00a} and non-unitary evolution\cite{yi03} from the
viewpoint of interferometry. For unitary evolution, the mixed
state geometric phase was defined as $\mbox{arg Tr}(U^{||}\rho)$
with $U^{||}$ representing the unitary parallel transport, whereas
in the case of non-unitary evolution, it was defined as a weighted
average over phase factors of the non-transition states. We will
adopt the latter definition to study the properties of geometric
phase for the subsystem. To start with, we write down the reduced
density matrix corresponding to the instantaneous eigenstate
$|\psi_1\rangle$ given by Eq.(\ref{eigens}) for subsystem 1 in
basis $\{|e_1,n\rangle, |g_1,n+1\rangle\}$ as
\begin{equation}
\rho_1=\left ( \matrix{ \cos^2\frac{\alpha}{2} &
\cos\frac{\alpha}{2}\sin\frac{\alpha}{2}e^{i\phi}\cr
\cos\frac{\alpha}{2}\sin\frac{\alpha}{2}e^{-i\phi} &
\sin^2\frac{\alpha}{2} } \right ),
\end{equation}
it yields the mixed state geometric phase for subsystem 1
\begin{equation}
\gamma_1^1=\pi(1+\cos\alpha), \label{mbp1}
\end{equation}
in the same way, the geometric phases for subsystem 1 pertaining
to the other instantaneous eigenstates $|\psi_2\rangle,
|\psi_3\rangle$ and $|\psi_4\rangle$ in Eq.(\ref{eigens}) can be
calculated  as follows
\begin{eqnarray}
\gamma_2^1&=&\pi(1-\cos\alpha),\nonumber\\
\gamma_3^1&=&\pi(1+\cos\beta),\nonumber\\
\gamma_4^1&=&\pi(1-\cos\beta),\label{mbp2}
\end{eqnarray}
These mixed state geometric phases are very similar to the
geometric phase of the single spin-$\frac 1 2 $ particle driving
by the classical magnetic field\cite{berry84}, aside from an
energy shift of $2J$ in $\cos\alpha$ and $\cos\beta$, the vacuum
induced correction to the geometric phases of subsystem 1 also
exist in this case, it can be seen from the definition of $\alpha$
and $\beta$ in Eq.(\ref{alpha}). The dependence of $\cos\alpha$
and $\cos\beta$ on the photon number $n$ and the spin-spin
coupling constant $J$ was represented in figure 1, where we choose
$\omega=\nu$, i.e., the spin-field coupling is on resonance to
plot the figure, $\cos\xi (\xi=\alpha, \beta)$ is the key quantity
in this study, since all these phases are related to this
quantity. From figure 1, we can see that the mixed state geometric
phases Eq. ( \ref{mbp1},\ref{mbp2}) tend to zero (or 2$\pi$) with
$J\rightarrow \infty$. The effect due to the field quantization is
also clear from Eq.(\ref{bpt1}); they tend to
$\frac{\pi}{2}\cos\theta$ with $J\rightarrow 0$ when the
field-particle interaction is on resonance, and these phases tend
to $\pi\cos\theta / 0$ in the strong spin-spin coupling limit
$J\rightarrow\infty$.

Finally, we discuss the problem of adiabaticity. In the standard
semiclassical theory, the driving field has to change slowly to
entail the system to undergo an adiabatic evolution, in the full
quantum regime the adiabatic condition follows straightforwardly
($p,q=1,2,3,4$) from
\begin{equation}
|\frac{\langle\psi_p(t)|\frac{\partial}{\partial
t}|\psi_q(t)\rangle}{E_p(t)-E_q(t)}|<<1, p\neq q,
\end{equation}
and it is given that $\frac{\omega}{4}\lambda\sqrt{n+1}<<1$, with
$\omega$ the precessing frequency ($\phi=\omega t$), which is
independent of the spin-spin couplings, this indicates that the
spin-spin couplings does not affect the adiabaticity of the
composite system, which is quite different from the case with
classical field driving \cite{yi042}. The extension of above
analysis to the case of system with intra-variable coupling is
straight forward. For example, consider an atom with electronic
orbital and spin angular momentum $\vec{L}$ and $\vec{s}$,
respectively, the spin-orbit coupling $\xi \vec{L}\cdot \vec{s}$
would play the role of the inter-subsystem coupling in the above
discussion. The atom in this case may be treated as a composite
system, and its geometric phase acquired when it transports round
a loop in the parameter space strongly depends on the spin-orbit
coupling, which is similar to the case with classical field
driving.

To sum up, we have calculated the pure state Berry phases of the
composite system and the mixed state geometric phases of its
subsystems. The results show that the inter-subsystem coupling
would diminish these phases and make the observation of the field
quantization effect difficult. However, this is not the case when
the driving field are of two modes, as Eq.(\ref{bpt1}) shows, the
Berry phases due to the vacuum tend to
$\frac{\pi}{2}(\cos\theta\pm \cos\theta)$ that would be a constant
with a specific $\theta$. The adiabatic condition does not depend
on the inter-subsystem coupling, this is quite different from the
case with classical field driving.

\vskip 0.3 cm XXY acknowledges financial support from EYTP of
M.O.E, and NSF of China under project 10305002.

\end{document}